\batchmode
\makeatletter
\def\input@path{{C:/Users/Zhentao/Dropbox/write/Rmosek/Report/}}
\makeatother
\documentclass[11pt,english]{article}\usepackage[]{graphicx}\usepackage[]{color}
\makeatletter
\def\maxwidth{ %
  \ifdim\Gin@nat@width>\linewidth
    \linewidth
  \else
    \Gin@nat@width
  \fi
}
\makeatother

\definecolor{fgcolor}{rgb}{0.345, 0.345, 0.345}
\newcommand{\hlnum}[1]{\textcolor[rgb]{0.686,0.059,0.569}{#1}}%
\newcommand{\hlstr}[1]{\textcolor[rgb]{0.192,0.494,0.8}{#1}}%
\newcommand{\hlcom}[1]{\textcolor[rgb]{0.678,0.584,0.686}{\textit{#1}}}%
\newcommand{\hlopt}[1]{\textcolor[rgb]{0,0,0}{#1}}%
\newcommand{\hlstd}[1]{\textcolor[rgb]{0.345,0.345,0.345}{#1}}%
\newcommand{\hlkwa}[1]{\textcolor[rgb]{0.161,0.373,0.58}{\textbf{#1}}}%
\newcommand{\hlkwb}[1]{\textcolor[rgb]{0.69,0.353,0.396}{#1}}%
\newcommand{\hlkwc}[1]{\textcolor[rgb]{0.333,0.667,0.333}{#1}}%
\newcommand{\hlkwd}[1]{\textcolor[rgb]{0.737,0.353,0.396}{\textbf{#1}}}%

\usepackage{framed}
\makeatletter
\newenvironment{kframe}{%
 \def\at@end@of@kframe{}%
 \ifinner\ifhmode%
  \def\at@end@of@kframe{\end{minipage}}%
  \begin{minipage}{\columnwidth}%
 \fi\fi%
 \def\FrameCommand##1{\hskip\@totalleftmargin \hskip-\fboxsep
 \colorbox{shadecolor}{##1}\hskip-\fboxsep
     \hskip-\linewidth \hskip-\@totalleftmargin \hskip\columnwidth}%
 \MakeFramed {\advance\hsize-\width
   \@totalleftmargin\z@ \linewidth\hsize
   \@setminipage}}%
 {\par\unskip\endMakeFramed%
 \at@end@of@kframe}
\makeatother

\definecolor{shadecolor}{rgb}{.97, .97, .97}
\definecolor{messagecolor}{rgb}{0, 0, 0}
\definecolor{warningcolor}{rgb}{1, 0, 1}
\definecolor{errorcolor}{rgb}{1, 0, 0}
\newenvironment{knitrout}{}{} 

\usepackage{alltt}
\usepackage[T1]{fontenc}
\usepackage[latin9]{inputenc}
\usepackage[letterpaper]{geometry}
\usepackage{color}
\usepackage{babel}
\usepackage{booktabs}
\usepackage{url}
\usepackage{amsmath}
\usepackage{amsthm}
\usepackage{amssymb}
\usepackage{graphicx}
\usepackage{setspace}
\usepackage[authoryear]{natbib}
\usepackage[
 bookmarks=true,bookmarksnumbered=false,bookmarksopen=false,
 breaklinks=false,pdfborder={0 0 0},pdfborderstyle={},backref=false,colorlinks=true]
 {hyperref}
\hypersetup{
 citecolor=blue}

\makeatletter

\providecommand{\tabularnewline}{\\}

\theoremstyle{definition}
 \newtheorem{example}{\protect\examplename}

\usepackage{listings}

\makeatother

\providecommand{\examplename}{Example}
\IfFileExists{upquote.sty}{\usepackage{upquote}}{}
\begin{document}
\title{Implementing Convex Optimization in R:\\ Two Econometric Examples}
\author{Zhan Gao \\ \small University of Southern California \vspace{5pt}\\  Zhentao Shi\\ \small The Chinese University of Hong Kong}

\date{}

\maketitle

\vspace{-.5cm}
\begin{abstract}
Economists specify high-dimensional models to address heterogeneity
in empirical studies with complex big data. Estimation of these models
calls for optimization techniques to handle a large number of parameters.
Convex problems can be effectively executed in modern statistical
programming languages. We complement Koenker and Mizera (2014)'s work
on numerical implementation of convex optimization, with focus on
high-dimensional econometric estimators. Combining \texttt{R} and
the convex solver \texttt{MOSEK} achieves faster speed and equivalent
accuracy, demonstrated by examples from Su, Shi, and Phillips (2016)
and Shi (2016). Robust performance of convex optimization is witnessed
cross platforms. The convenience and reliability of convex optimization
in \texttt{R} make it easy to turn new ideas into prototypes. 
\end{abstract}
\vspace{1cm}



\vspace{1cm}

\small \noindent Zhan Gao: \texttt{zhangao@usc.edu}. Address: Department
of Economics, University of Southern California, 3620 South Vermont
Ave. Kaprielian Hall, 300 Los Angeles, CA 90089-0253, USA. Zhentao
Shi (corresponding author): \texttt{zhentao.shi@cuhk.edu.hk}. Address:
Department of Economics, 912 Esther Lee Building, the Chinese University
of Hong Kong, Sha Tin, New Territories, Hong Kong SAR, China. Tel:
(852) 3943-1432. Fax (852) 2603-5805. We thank Roger Koenker for inspiration
and hospitality during the second author's visit to University of
Illinois.

\newpage{}

\normalsize

\section{Introduction}

Equipped with tremendous growth of computing power over the last few
decades, econometricians endeavor to tackle high-dimensional real
world problems that we could hardly have imagined before. Along with
the development of modern asymptotic theory, computation has gradually
ascended onto the central stage. Today, discussion of numerical algorithms
is essential for new econometric procedures. 

Optimization is at the heart of estimation, and convex optimization
is the best understood category. Convex problems are ubiquitous in
econometric textbooks. The least square problem is convex, and the
classical normal regression is also convex after straightforward reparametrization.
Given a linear single-index form, the Logit or Probit binary regression,
the Poisson regression and the regressions with a censored or truncated
normal distributions are all convex. Another prominent example is
the quantile regression \citep{koenker1978regression}, motivated
from its robustness to non-Gaussian errors and outlier contamination.

With the advent of big data, practitioners attempt to build general
models that involve hundreds or even more parameters in the hope to
capture complex heterogeneity in empirical economic studies. Convex
optimization techniques lay out the foundation of estimating these
high-dimensional models. Recent years witnesses \citet{bajari2015machine},
\citet{gu2017empirical} and \citet{doudchenko2016balancing}, to
name a few, exploring new territories by taking advantage of convexity. 

To facilitate practical implementation, \citet{cvxinr} summarize
the operation in \texttt{R} by \texttt{MOSEK} via \texttt{Rmosek}
to solve linear programming, conic quadratic programming, quadratic
programming, etc. \texttt{R} is open-source software, \texttt{MOSEK}
is a proprietary convex optimization solver but offers free academic
license, and \texttt{Rmosek} is the \texttt{R} interface that communicates
with \texttt{MOSEK}. \texttt{MOSEK} specializes in convex problems
with reliable performance, and is  competitive in high-dimensional
problems.

This paper complements \citet{cvxinr}'s work. We replicate by \texttt{Rmosek}
two examples of high-dimensional estimators, namely \citet{su2016identifying}'s
classifier-Lasso (C-Lasso) and \citet{REL}'s relaxed empirical likelihood
(REL). In addition, we replicate by \texttt{Rmosek} an application
of C-Lasso that re-examines China's GDP growth rate \citep{chinagdp}.
These exercises highlight two points. Firstly, the \texttt{R} environment
is robust in numerical accuracy for high-dimensional convex optimization
and \texttt{Rmosek} takes the lead in computational speed. Second,
we showcase the ease of creating new econometric estimators\textemdash often
no more than a few lines of code\textemdash by the code snippets in
the Appendix. Such convenience lowers the cost of turning an idea
into a prototype, and enables researchers to glean valuable insights
about their archetypes by experimenting new possibilities. All code
in this note is hosted at \url{https://github.com/zhentaoshi/convex_prog_in_econometrics}.

\section{Classifier-Lasso \label{sec:Classifier-Lasso}}

It is common practice to assume in linear fixed-effect panel data
models that the cross-sectional units are heterogeneous in terms of
the time-invariant individual intercept, while they all share the
same slope coefficient. This pooling assumption can be tested and
is often rejected in real-world applications. In recent years panel
data group structure has been attracting attention. \citet{bonhomme2015grouped}
allow group structure in the intercept and use the $k$-means algorithm
for classification. When the slope coefficients exhibit group structure,
\citet{su2016identifying} propose Classifier-Lasso (C-Lasso) to identify
the latent group pattern. 

We illustrate the penalized least square (PLS), a simple special case
of C-Lasso.\footnote{The profile log-likelihood function $Q_{1,nT}\left(\beta\right)=\sum_{i=1}^{n}\sum_{t=1}^{T}\psi\left(w_{it},\beta_{i},\hat{\mu}_{i}\left(\beta_{i}\right)\right)$
for nonlinear models can be reformulated into a separable form, while
penalized GMM (PGMM) can be handled under the same optimization framework
as PLS. They are discussed in Appendix \ref{subsec:Nonlinear-Lasso}
and Appendix \ref{subsec:Penalized-GMM}, respectively.} Given a tuning parameter $\lambda$ and the number of groups $K$,
PLS is defined as the solution to
\[
\min_{\boldsymbol{\beta},\left(\alpha_{k}\right)_{k=1}^{K}}\,\frac{1}{nT}\sum_{i=1}^{n}\sum_{t=1}^{T}\left(y_{it}-x_{it}^{\prime}\beta_{i}\right)^{2}+\frac{\lambda}{n}\sum_{i=1}^{n}\prod_{k=1}^{K}\left\Vert \beta_{i}-\alpha_{k}\right\Vert _{2}
\]
where $\boldsymbol{\beta}=\left(\beta_{i}\right)_{i=1}^{n}$. The
additive-multiplicative penalty pushes the individual slope coefficients
$\beta_{i}$ in the same group toward a common coefficient $\alpha_{k}$.
This is not a convex problem, but the optimization with the additive-multiplicative
penalty can be approximated by an iterative algorithm, as is explained
in the Supplement of \citet[Section S3.1]{su2016identifying}. Procedures
based on such an iteration have been successfully applied to \citet{su2017interactive},
\citet{su2017numberofgroup} and \citet{su2017sieve}. The iterative
algorithm initiates at the within-group estimator, which is consistent
when $T$ is large. In the $k$-th sub-step of the $r$-th iteration,
$(\boldsymbol{\beta},\alpha_{\tilde{k}})$ is chosen to minimize
\begin{equation}
\min_{\boldsymbol{\beta},\alpha_{\tilde{k}}}\,\frac{1}{nT}\sum_{i=1}^{n}\sum_{t=1}^{T}\left(y_{it}-x_{it}^{\prime}\beta_{i}\right)^{2}+\frac{\lambda}{n}\sum_{i=1}^{n}\left\Vert \beta_{i}-\alpha_{\tilde{k}}\right\Vert _{2}\gamma_{i}\label{eq:iter}
\end{equation}
where $\gamma_{i}=\prod_{k=1}^{\tilde{k}-1}\Vert\hat{\beta}_{i}^{(r,k)}-\hat{\alpha}_{k}^{(r)}\Vert_{2}\cdot\prod_{k=\tilde{k}+1}^{K}\Vert\hat{\beta}_{i}^{(r-1,k)}-\hat{\alpha}_{k}^{(r-1)}\Vert_{2}.$
The iteration proceeds until the $K$-convex problem numerically converges. 

Given the multiplier $\gamma_{i}$, the above optimization problem
is convex in $(\boldsymbol{\beta},\alpha_{\tilde{k}})$ and the structure
is very close to Lasso. Though the \texttt{R} packages \texttt{lars}
and\texttt{ glmnet} packages can carry out the standard Lasso, however,
it is not straightforward how to modify these functions to accommodate
(\ref{eq:iter}), where $\alpha_{\tilde{k}}$ is also an unknown parameter
to be optimized. A quick review of \citet{cvxinr} approach to Lasso
will be helpful.

\bigskip
\begin{example}[Lasso]
 The standard Lasso problem is 
\[
\min_{\beta}\,\frac{1}{n}\left\Vert y-X\beta\right\Vert _{2}^{2}+\lambda\left\Vert \beta\right\Vert _{1}
\]
where $y\in\mathbb{R}^{n}$ and $X\in\mathbb{R}^{n\times p}$ are
observed data, $\lambda$ is the tuning parameter and $\beta\in\mathbb{R}^{p}$
is the parameter of interests. However, \texttt{Rmosek} does not  accept
the $l_{1}$ norm. To overcome the difficulty, \citet{cvxinr} introduce
new parameters to transform the $l_{1}$-penalized problem into a
conic optimization that \texttt{Rmosek} recognizes. We first deal
with $\left\Vert \beta\right\Vert _{1}$. The $p\times1$ vector $\beta$
can be decomposed into a positive part $\beta^{+}=\left(\max\left\{ 0,\beta_{j}\right\} \right)_{j=1}^{p}$
and a negative part $\beta^{-}=\left(\max\left\{ 0,-\beta_{j}\right\} \right)_{j=1}^{p}$,
so that $\beta=\beta^{+}-\beta^{-}$ and $\left\Vert \beta\right\Vert _{1}=e^{\prime}\beta^{+}+e^{\prime}\beta^{-}$,
where $e$ is the $p\times1$ vector with all elements equal to $1$.
Next, we transform the $l_{2}$-norm $\left\Vert y-X\beta\right\Vert _{2}^{2}$
to a second-order conic constraint. Consider a minimization problem
with $\left\Vert v\right\Vert _{2}^{2}$ in the objective function.
We can use a new parameter $t$ to replace it and add a conic constraint
$\left\Vert v\right\Vert _{2}^{2}\leq t$, which is equivalent to
$\left\Vert \left(v,\frac{t-1}{2}\right)\right\Vert _{2}\leq\frac{t+1}{2}$.
Thus we obtain a standard conic constraint $\left\Vert \left(v,s\right)\right\Vert _{2}\leq r$,
where $s=\frac{t-1}{2}$ and $r=\frac{t+1}{2}$. We rewrite the Lasso
problem as
\begin{gather*}
\min_{\theta}\,\lambda\left(e'\beta^{+}+e'\beta^{-}\right)+\frac{t}{n}\\
\text{s.t.\ \ }v=y-X\left(\beta^{+}-\beta^{-}\right),\,\left\Vert \left(v,s\right)\right\Vert _{2}\leq r,\,s=\frac{t-1}{2},\,r=\frac{t+1}{2}
\end{gather*}
where $\theta=\left(\beta^{+},\beta^{-},v,t,s,r\right)$. This problem
is of the standard form of second-order conic programming and hence
can be executed in \texttt{Rmosek}.

\bigskip
\end{example}
Applying the techniques in the Lasso formulation, we can transform
the $l_{2}$-norm terms in (\ref{eq:iter}) and formulate the problem
into a conic programming:
\begin{gather*}
\min_{\alpha_{\tilde{k}},\theta}\,\sum_{i=1}^{n}\left(\left(\frac{1}{nT}\right)t_{i}+\left(\frac{\lambda}{n}\gamma_{i}\right)w_{i}\right)\\
\text{s.t.}\;\;x_{i}\beta_{i}+\nu_{i}=y_{i},\,\beta_{i}-\mu_{i}-\alpha_{\tilde{k}}=0,\,s_{i}-\frac{1}{2}t_{i}=-\frac{1}{2},\:r_{i}-\frac{1}{2}t_{i}=\frac{1}{2},\\
\left\Vert \left(\nu_{i},s_{i}\right)\right\Vert _{2}\leq r_{i},\,\left\Vert \mu_{i}\right\Vert _{2}\leq w_{i},\,t_{i}\geq0,\,\text{for all}\,i=1,2,\cdots,n
\end{gather*}
where $\theta=\left\{ \beta_{i},\nu_{i},\mu_{i},s_{i},r_{i},t_{i},w_{i}\right\} _{i=1}^{n}$.
The convexity is manifest when we write the problem in matrix form,
as is displayed in Appendix \ref{sec:Code-Snippets}. 

\subsection{Replication}

We replicate\textcolor{black}{{} }the simulation studies in \citet[Section 4]{su2016identifying}
in \texttt{R} via \texttt{Rmosek} and compare the performance of different
numerical optimization approaches. \citet{su2016identifying} conduct
their numerical work in \texttt{MATLAB} via \texttt{CVX} \citep{cvx}.
\texttt{CVX} is a \texttt{MATLAB} add-on package for \emph{disciplined
convex optimization} \citep[DCP]{grant2006disciplined}. It provides
an interface to communicate with commercial or open-source solvers.
In the \texttt{R }\textcolor{black}{environment, the de facto solver
is }\texttt{\textcolor{black}{optimx}}\textcolor{black}{{} \citep{optimx};
another option is the interface }\texttt{\textcolor{black}{nloptr}}\textcolor{black}{{}
\citep{nloptr} that hooks optimization solver }\texttt{\textcolor{black}{NLopt}}\textcolor{black}{{}
\citep{nlopt}. They are general-purpose optimization solvers not
tailored for convexity. Most recently, \citet{CVXR} are actively
developing }\texttt{\textcolor{black}{CVXR}}\textcolor{black}{, }\texttt{\textcolor{black}{CVX}}\textcolor{black}{'s
counterpart in }\texttt{\textcolor{black}{R}}\textcolor{black}{. At
this stage, it is integrated with the open-source solver }\texttt{\textcolor{black}{ECOS}}\textcolor{black}{{}
\citep{ECOS}. We also consider the counterpart of }\texttt{\textcolor{black}{CVX}}\textcolor{black}{{}
in }\texttt{\textcolor{black}{Python}}\textcolor{black}{{} environment,
}\texttt{\textcolor{black}{CVXPY}}\textcolor{black}{{} \citep{cvxpy},
to verify the stability of the algorithm across platforms.}\footnote{\textcolor{black}{In the latest version (Version 0.99), }\texttt{\textcolor{black}{CVXR}}\textcolor{black}{{}
supports }\texttt{\textcolor{black}{MOSEK}}\textcolor{black}{{} by sending
the problem to }\texttt{\textcolor{black}{MOSEK}}\textcolor{black}{{}
in the }\texttt{\textcolor{black}{Python}}\textcolor{black}{{} environment.
In our experiment, large-scale problems like the C-Lasso cause errors
in the communication between }\texttt{\textcolor{black}{R}}\textcolor{black}{{}
and }\texttt{\textcolor{black}{Python}}\textcolor{black}{. In addition,
}\texttt{\textcolor{black}{CVXR}}\textcolor{black}{{} with }\texttt{\textcolor{black}{MOSEK}}\textcolor{black}{{}
currently cannot incorporate problems with nonlinear objective functions
and hence cannot be used for REL in Section \ref{sec:Relaxed-Empirical-Likelihood}. }}

\begin{table}[h]
\caption{\label{tab:classopls}Classification and Point Estimation of $\alpha_{1}$:
Replication of \citet[DGP 1]{su2016identifying}}

\medskip{}

\centering{}%
\begin{tabular}{rrrrrrr}
\toprule 
$\left(n,T\right)$ &
$\left(100,15\right)$ &
$\left(100,25\right)$ &
$\left(100,50\right)$ &
$\left(200,15\right)$ &
$\left(200,25\right)$ &
$\left(200,50\right)$\tabularnewline
\midrule 
\multicolumn{7}{c}{RMSE}\tabularnewline
\texttt{Rmosek} &
0.0762 &
0.0386 &
0.0247 &
0.0428 &
0.0278 &
0.0174\tabularnewline
\texttt{CVXR} &
0.0762 &
0.0386 &
0.0247 &
0.0427 &
0.0278 &
0.0174\tabularnewline
\texttt{CVX} &
0.0767 &
0.0399 &
0.0253 &
0.0443 &
0.0286 &
0.0179\tabularnewline
\texttt{CVXPY} &
0.0741 &
0.0394 &
0.0253 &
0.0424 &
0.0271 &
0.0173\tabularnewline
\midrule
\multicolumn{7}{c}{Correct Ratio}\tabularnewline
\texttt{Rmosek} &
0.8987 &
0.9645 &
0.9965 &
0.9019 &
0.9668 &
0.9969\tabularnewline
\texttt{CVXR} &
0.8986 &
0.9645 &
0.9965 &
0.9020 &
0.9668 &
0.9969\tabularnewline
\texttt{CVX} &
0.8991 &
0.9647 &
0.9965 &
0.9026 &
0.9667 &
0.9968\tabularnewline
\texttt{CVXPY} &
0.8988 &
0.9644 &
0.9965 &
0.9021 &
0.9667 &
0.9969\tabularnewline
\midrule
\multicolumn{7}{c}{Running Time (in minute) }\tabularnewline
\texttt{Rmosek} &
18.08 &
10.58 &
8.42 &
24.94 &
15.87 &
13.57\tabularnewline
\texttt{CVXR} &
73.85 &
40.75 &
32.51 &
77.86 &
47.26 &
37.03\tabularnewline
\texttt{CVX} &
94.57 &
50.25 &
34.98 &
90.14 &
54.29 &
41.27\tabularnewline
\texttt{CVXPY} &
27.91 &
18.17 &
26.32 &
33.64 &
26.70 &
29.09\tabularnewline
\bottomrule
\end{tabular}
\end{table}

We follow DGP 1 in \citet[Section 4]{su2016identifying} as a benchmark.
Table \ref{tab:classopls} reports under various combinations of the
cross sectional units $n$ and the time length $T$, the root-mean-square
error (RMSE) of $\widehat{\alpha}_{1}$ and the probability of correct
group classification (correct ratio). The DGP, simulation settings
and the indicators are relegated to Appendix \ref{subsec:C-Lasso DGP}
to save space.

Within the \texttt{R} environment, the numerical results of estimation
error and classification correct ratio by \texttt{Rmosek} are almost
identical to \texttt{CVXR} up to rounding errors.\footnote{The de facto solver \texttt{optimx} breaks down when solving such
high dimensional problems. \texttt{nloptr} takes more than a few hours
to finish one estimation, which makes the full-scale simulation exercise
computational infeasible. In addition, \texttt{nloptr} fails to attain
an accurate solution in most cases according to our experiments.} We also implement the simulation in \texttt{MATLAB} via \texttt{CVX}
and in \texttt{Python} via \texttt{CVXPY}, the results are largely
similar, which demonstrates the robustness of the numerical performance
of C-Lasso across different computing platforms.

Practitioners may need to try out different specifications for robustness
check in real applications. Without fast optimization solvers, computational
cost can become a bottleneck. On the same computing platform of Intel(R)
Core(TM) i7-8750H CPU @ 2.20GHz, each case is executed in a single
thread and we record the running time in the lower panel in Table
\ref{tab:classopls}. \texttt{Rmosek} significantly outperforms all
alternatives. \texttt{CVX} in \texttt{MATLAB} is about $3$ to $5.2$
times slower and \texttt{CVXPY} is about $1.4$ to $3.1$ times slower
than \texttt{Rmosek}. Although \texttt{CVX} and \texttt{CVXPY} are
also powered by \texttt{MOSEK}, the DCP system takes time to check
the convexity of the input problem and automate the formulation. For
similar reasons, \texttt{CVXR} is $2.7$ to $4$ times slower than
\texttt{Rmosek}. According to \citet[Section 4.1]{CVXR}, we can skip
the DCP formulation steps with \texttt{CVXR} and the advantage of
\texttt{Rmosek} becomes around $1.5$ to $1.8$, which illustrates
the advantages of \texttt{MOSEK} over the open-source solver \texttt{ECOS}.
In summary, DCP is useful when we are uncertain about the convexity
and solvability of a problem. However, for problems that are mathematically
verified to be convex, directly calling \texttt{MOSEK} saves much
computational time.

\subsection{Empirical Application\label{subsec:Empirical-Application}}

While China is now the second largest economy in the world in terms
of aggregate GDP, the accuracy of its reported national income accounting
has been a topic of constant debate over the years. Most recently,
\citet{chinagdp} utilize local economic indicators that are directly
associated with economic activities to estimate China's local and
aggregate GDP in order to assess the quality of these numbers. Different
regions of this continent-size country are growing at varying pace,
thereby resulting in tremendous heterogeneity among its provinces.
To control the hidden heterogeneity, \citet{chinagdp} specify a linear
fixed effect model with latent group structure
\[
y_{it}=x_{it}^{\prime}\beta_{i}+v_{i}+\varepsilon_{it},
\]
where $y_{it}$ is the logarithm of GDP for province $i$ at year
$t$, $x_{it}$ includes local economic indicators of interests, $v_{i}$
characterizes the fixed effect of province $i$, and $\varepsilon_{it}$
is the idiosyncratic error. The heterogeneous slope coefficients $\beta_{i}$
captures latent group structures across regions to be determined by
C-Lasso. The data span from year 2000 to 2007, i.e. $T=8$. Five indicators
are employed as regressors to control observable heterogeneity, namely
\emph{satellite night lights}, \emph{national tax revenue}, \emph{exports},
\emph{imports}, and \emph{electricity consumption}. These indicators
are less susceptible to local officials' manipulation and thus more
robustly reflect economic activities for real businesses. Two alternative
specifications, one with no \emph{satellite night lights} and the
other with neither \emph{satellite night lights} nor \emph{national
tax revenue}, are also considered.

In our implementation, the number of groups $K$ and tuning parameter
$\lambda$ are determined by the information criterion proposed in
\citet[Section 2.5]{su2016identifying}. Trials with different specifications
and tuning parameters can be time-consuming, particularly for our
panel data of a short $T$ as the algorithm has to iterate many times
until numerical convergence. \citet{chinagdp} estimate the model
in \texttt{MATLAB} via \texttt{CVX} and the classification results
are reported in \citet[Table A11]{chinagdp}. We replicate the classification
results by \texttt{Rmosek} and compare the accuracy and speed to \texttt{CVX}.

\begin{table}
\caption{\label{tab:chinagdp_result_table}Classification Results: Replication
of \citet[Table A11]{chinagdp}}

\medskip{}

\begin{centering}
\begin{tabular}{llllllll}
\toprule 
\multicolumn{2}{c}{All 5 Indicators} &
 &
\multicolumn{2}{c}{Without Light} &
 &
\multicolumn{2}{c}{Without Light and Tax}\tabularnewline
Group 1 &
Group 2 &
 &
Group 1 &
Group 2 &
 &
Group 1 &
Group 2\tabularnewline
\midrule
{\small{}Beijing} &
{\small{}Tianjin} &
 &
{\small{}Beijing} &
{\small{}Tianjin} &
 &
{\small{}Beijing} &
{\small{}Tianjin}\tabularnewline
{\small{}Hebei} &
{\small{}Jilin} &
 &
{\small{}Inner Mongolia} &
{\small{}Hebei} &
 &
{\small{}Hebei} &
{\small{}Liaoning}\tabularnewline
{\small{}Shanxi} &
{\small{}Heilongjiang} &
 &
{\small{}Liaoning} &
{\small{}Shanxi} &
 &
{\small{}Shanxi} &
{\small{}Shanghai}\tabularnewline
{\small{}Inner Mongolia} &
\textbf{\small{}Liaoning} &
 &
{\small{}Jilin} &
{\small{}Heilongjiang} &
 &
{\small{}Inner Mongolia} &
{\small{}Zhejiang}\tabularnewline
{\small{}Shanhai} &
{\small{}Jiangxi} &
 &
{\small{}Shanghai} &
{\small{}Zhejiang} &
 &
{\small{}Jilin} &
{\small{}Shandong}\tabularnewline
{\small{}Jiangsu} &
{\small{}Henan} &
 &
{\small{}Jiangsu} &
{\small{}Jiangxi} &
 &
{\small{}Heilongjiang} &
{\small{}Henan}\tabularnewline
{\small{}Anhui} &
{\small{}Hunan} &
 &
{\small{}Anhui} &
{\small{}Shandong} &
 &
{\small{}Jiangsu} &
{\small{}Hunan}\tabularnewline
{\small{}Fujing} &
{\small{}Guangdong} &
 &
{\small{}Fujian} &
{\small{}Henan} &
 &
{\small{}Anhui} &
{\small{}Guangdong}\tabularnewline
{\small{}Hubei} &
\textbf{\small{}Zhejiang} &
 &
{\small{}Hubei} &
{\small{}Hunan} &
 &
{\small{}Fujian} &
{\small{}Chongqing}\tabularnewline
{\small{}Hainan} &
{\small{}Guangxi} &
 &
{\small{}Hainan} &
{\small{}Guangdong} &
 &
{\small{}Jiangxi} &
{\small{}Guizhou}\tabularnewline
{\small{}Qinghai} &
{\small{}Chongqing} &
 &
{\small{}Qinghai} &
{\small{}Guangxi} &
 &
{\small{}Hubei} &
{\small{}Yunnan}\tabularnewline
{\small{}Xinjiang} &
{\small{}Sichuan} &
 &
 &
{\small{}Chongqing} &
 &
{\small{}Guangxi} &
{\small{}Shaanxi}\tabularnewline
 &
\textbf{\small{}Shandong} &
 &
 &
{\small{}Sichuan} &
 &
{\small{}Hainan} &
\tabularnewline
 &
{\small{}Guizhou} &
 &
 &
{\small{}Guizhou} &
 &
{\small{}Sichuan} &
\tabularnewline
 &
{\small{}Shaanxi} &
 &
 &
{\small{}Yunnan} &
 &
{\small{}Gansu} &
\tabularnewline
 &
{\small{}Gansu} &
 &
 &
{\small{}Shaanxi} &
 &
{\small{}Qinghai} &
\tabularnewline
 &
\textbf{\small{}Yunnan} &
 &
 &
{\small{}Gansu} &
 &
{\small{}Ningxia} &
\tabularnewline
 &
{\small{}Ningxia} &
 &
 &
{\small{}Ningxia} &
 &
{\small{}Xinjiang} &
\tabularnewline
 &
 &
 &
 &
{\small{}Xinjiang} &
 &
 &
\tabularnewline
\bottomrule
\end{tabular}
\par\end{centering}
\medskip{}

\raggedright{}\small Note: Provinces in bold highlight different
results from the original paper.
\end{table}

The same as in the original paper, in all three specifications the
information criterion determines two groups. Displayed in Table \ref{tab:chinagdp_result_table},
the classification results by \texttt{Rmosek} are identical to those
by \texttt{CVX} reported in \citet[Table A11]{chinagdp} in the second
and the third specifications. However, in the first specification
where all five indicators are included, we observe deviance across
computing platforms. Liaoning, Zhejiang, Shandong and Yunnan are moved
into group 2 according to \texttt{Rmosek} whereas they are left in
the group 1 in \texttt{CVX} results. Since this is a short $T$ panel
data with $T=8$, the numerical stability is more fragile when we
include more regressors.

\begin{figure}
\caption{\label{fig:Tertiary-industry-GDP} GDP Shares of Tertiary Sector}

\medskip{}

\centering{}\includegraphics[width=0.9\textwidth]{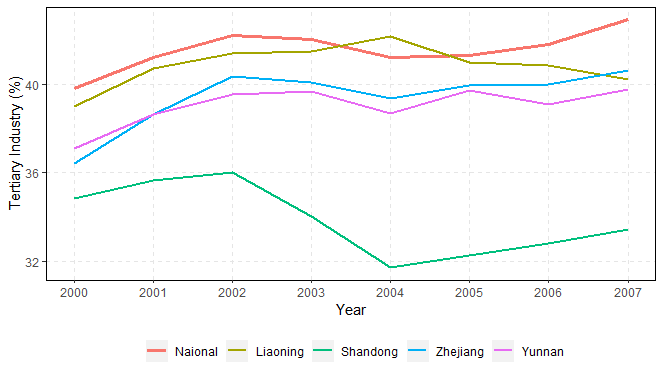}
\end{figure}

A key observation in \citet{chinagdp} is that group 1 gathers Beijing,
Shanghai and Hainan, the three provinces with the highest GDP shares
of the tertiary sector associated with the provision of services.
The \texttt{Rmosek} result retains this feature. As shown in Figure
\ref{fig:Tertiary-industry-GDP}, the tertiary industry GDP shares
of the four provinces excluded from the Beijing-Shanghai-Hainan group
doe not reach the national aggregate share, except for Liaoning in
a single year 2004. Given their relatively lower shares, it is sensible
that \texttt{Rmosek} removes them out of the high-share group.

In the parameter tuning process, we compute the information criterion
for $K=1,2,3,4$ and 10 candidate $\lambda$ values.\footnote{We generate $\lambda=c\mathrm{var}\left(y\right)T^{-\frac{1}{3}}$
where the constant $c$ varies from $0.001$ to $0.01$.} We report the CPU time consumed by the parameter tuning process in
Table \ref{tab:Running-Time-china-gdp}. \texttt{Rmosek} is about
$8$ to $9.6$ times faster than \texttt{CVX}. It demonstrates the
speed gain of \texttt{Rmosek} in real applications.

\begin{table}

\caption{\label{tab:Running-Time-china-gdp}Running Time (in second): Replication
of \citet[Table A11]{chinagdp}}

\medskip{}

\begin{centering}
\begin{tabular}{cccc}
\toprule 
 &
{\small{}With Light} &
{\small{}Without Light} &
{\small{}Without Light and Tax}\tabularnewline
\midrule
\texttt{Rmosek} &
$72.39$ &
$96.37$ &
$40.41$\tabularnewline
\texttt{CVX} &
$692.23$ &
$774.96$ &
$337.10$\tabularnewline
\bottomrule
\end{tabular}
\par\end{centering}
\end{table}

\section{Relaxed Empirical Likelihood \label{sec:Relaxed-Empirical-Likelihood}}

Besides the regression setting in Section \ref{sec:Classifier-Lasso},
convex programming is also useful in structural econometric estimation.
Consider the models with a ``true'' parameter $\beta_{0}$ satisfying
the unconditional moment condition $\mathbb{E}\left[g\left(Z_{i},\beta_{0}\right)\right]=\boldsymbol{0}_{m}$,
where $\left\{ Z_{i}\right\} _{i=1}^{n}$ is the observed data, $\beta\in\mathcal{B\subset}\mathbb{R}^{D}$
is a finite dimensional vector in the parameter space $\mathcal{B}$,
and $g$ is an $\mathbb{R}^{m}$-valued moment function. GMM \citep{hansen1982GMM}
and empirical likelihood (EL) \citep{owen1988empirical,qin1994empirical}
are two workhorses dealing with moment restriction models. In particular,
EL solves 
\[
\max_{\beta\in\mathcal{B},\pi\in\Delta_{n}}\,\sum_{i=1}^{n}\log\pi_{i}\quad\text{s.t.}\quad\sum_{i=1}^{n}\pi_{i}g\left(Z_{i},\beta\right)=\boldsymbol{0}_{m}
\]
where $\Delta_{n}=\left\{ \pi\in\left[0,1\right]^{n}:\sum_{i=1}^{n}\pi_{i}=1\right\} $
is the $n$-dimensional probability simplex. However, neither GMM
nor EL can be used to estimate a model with more moment equalities
than observations, i.e. $m>n$. To make the optimization feasible,
\citet{REL} relaxes the equality restriction $\sum_{i=1}^{n}\pi_{i}g_{i}\left(\beta\right)=\boldsymbol{0}_{m}$
in EL. REL is defined as the solution to 
\[
\max_{\beta\in\mathcal{B}}\max_{\pi\in\Delta_{n}^{\lambda}\left(\beta\right)}\,\sum_{i=1}^{n}\log\pi_{i}
\]
where 
\[
\Delta_{n}^{\lambda}\left(\beta\right)=\left\{ \pi\in\Delta_{n}:\big|\sum_{i=1}^{n}\pi_{i}h_{ij}\left(\beta\right)\big|\leq\lambda,\:j=1,2,\cdots,m\right\} 
\]
is a relaxed simplex, $\lambda\geq0$ is a tuning parameter, $h_{ij}\left(\beta\right)=g_{j}\left(Z_{i},\beta\right)/\hat{\text{\ensuremath{\sigma}}}_{j}\left(\beta\right)$,
$g_{j}\left(Z_{i},\beta\right)$ is the $j$-th component of $g\left(Z_{i},\beta\right)$,
and $\hat{\text{\ensuremath{\sigma}}}_{j}\left(\beta\right)$ is the
sample standard deviation of $\left\{ g_{j}\left(Z_{i},\beta\right)\right\} _{i=1}^{n}$.
The formulation of REL is inspired by Dantzig selector \citep{dantzig}. 

\bigskip
\begin{example}[Dantzig selector]
 Similar to Lasso, Dantzig selector also produces a sparse solution
to the linear regression model. Dantzig selector can be written as
\[
\min_{\beta}\,\lVert\beta\rVert_{1}\ \ \text{s.t.}\ \ \lVert X'\left(y-X\beta\right)\rVert_{\infty}\leq\lambda,
\]
where $\lambda$ is a tuning parameter. We can immediately reformulate
it as a linear programming problem 
\begin{gather*}
\min_{\beta^{+},\beta^{-}}\:e'\beta^{+}+e'\beta^{-}\\
\text{s.t.}\ \ X'y-\lambda e\leq\left(X'X\right)\left(\beta^{+}-\beta^{-}\right)\leq X'y+\lambda e\\
\beta^{+},\beta^{-}\geq0.
\end{gather*}
It is readily solvable using the \texttt{R} package \texttt{quantreg}
\citep{quantreg}. 
\end{example}
\bigskip 

Dantzig selector slacks the sup-norm of the first-order condition
for optimality. REL borrows the idea to estimate a finite-dimensional
parameter in a structural economic model defined by many moment equalities.
Comparing to Dantzig selector, REL uses a nonlinear objective function.
It is still convex (in minus likelihood) but \texttt{quantreg} that
deals with linear programming problems is no longer applicable. 

Similar to standard EL, REL's optimization involves an inner loop
and an outer loop. The outer loop for $\beta$ is a general low-dimensional
nonlinear optimization, which can be solved by Newton-type methods.
With the linear constraints and the logarithm objective, the inner
loop is convex in $\pi=\left(\pi_{i}\right)_{i=1}^{n}$. For each
$\beta$, the inner problem can be formulated as a \emph{separable
convex optimization problem} in the matrix form 
\begin{gather*}
\max_{\pi}\,\sum_{i=1}^{n}\log\pi_{i}\\
\text{s.t.}\ \ \begin{bmatrix}1\\
-\lambda\\
\vdots\\
-\lambda
\end{bmatrix}\leq\begin{bmatrix}1 & 1 & \cdots & 1\\
h_{11}\left(\beta\right) & h_{21}\left(\beta\right) & \cdots & h_{n1}\left(\beta\right)\\
\vdots & \vdots & \ddots & \vdots\\
h_{1m}\left(\beta\right) & h_{2m}\left(\beta\right) & \cdots & h_{nm}\left(\beta\right)
\end{bmatrix}\begin{bmatrix}\pi_{1}\\
\pi_{2}\\
\vdots\\
\pi_{n}
\end{bmatrix}\leq\begin{bmatrix}1\\
\lambda\\
\vdots\\
\lambda
\end{bmatrix}\\
0\leq\pi_{i}\leq1,\:\text{for each}\,i=1,2,\cdots,n
\end{gather*}
and it is readily solvable in \texttt{Rmosek} by translating the mathematical
expression into computer code.

\subsection{Replication}

We follow the simulation design in \citet[Section 4]{REL}, which
is described in Appendix \ref{subsec:REL-DGP}. Table \ref{tab:relsimulation}
reports the bias and RMSE of the estimation of $\hat{\beta}_{1}$,
implemented purely in \texttt{R} with the inner loop by \texttt{Rmosek}
and the outer loop by \texttt{nloptr}. The results are close to those
in \citet{REL}, where the code is written in \texttt{MATLAB} with
the outer loop handled by the function \texttt{fmincon} and the inner
loop by \texttt{CVX} solved by \texttt{MOSEK}. 

\begin{table}[h]
\caption{\label{tab:relsimulation}Estimation of $\beta_{1}$ in linear IV
model with REL: Replication of \citet{REL} }

\medskip{}

\centering{}%
\begin{tabular}{rcccccc}
\toprule 
 &
 &
\multicolumn{2}{c}{Replication} &
 &
\multicolumn{2}{c}{Original Results}\tabularnewline
\cmidrule{3-4} \cmidrule{4-4} \cmidrule{6-7} \cmidrule{7-7} 
$\left(n,m\right)$ &
 &
Bias &
RMSE &
 &
Bias &
RMSE\tabularnewline
\midrule
$\left(120,80\right)$ &
 &
-0.020 &
0.135 &
 &
-0.004 &
0.113\tabularnewline
$\left(120,160\right)$ &
 &
-0.018 &
0.162 &
 &
-0.012 &
0.143\tabularnewline
$\left(240,80\right)$ &
 &
-0.004 &
0.078 &
 &
-0.006 &
0.071\tabularnewline
$\left(240,160\right)$ &
 &
-0.008 &
0.093 &
 &
-0.009 &
0.077\tabularnewline
\bottomrule
\end{tabular}
\end{table}

We also experiment with other numerical alternatives. Since the scale
of the optimization problems here is much smaller than C-Lasso, the
inner loop can be correctly solved by \texttt{Rmosek}, \texttt{CVXR},
\texttt{CVX} in \texttt{MATLAB}, or even \texttt{nloptr}. These four
methods produce virtually identical inner loop results up to rounding
errors. This finding confirms the robustness of the \texttt{R} environment
in high-dimensional optimization. The difference in Table \ref{tab:relsimulation},
therefore, is attributed to the outer loop between the function \texttt{nloptr}
in \texttt{R} and the function \texttt{fmincon} in  \texttt{MATLAB}.

\begin{table}
\caption{\label{tab:rel_running_time}Running time of REL's inner loop \textcolor{black}{(in
second)}}

\medskip{}

\centering{}%
\begin{tabular}{crrrr}
\toprule 
$\left(n,m\right)$ &
$\left(120,80\right)$ &
$\left(120,160\right)$ &
$\left(240,80\right)$ &
$\left(240,160\right)$\tabularnewline
\midrule
\texttt{Rmosek} &
2.995 &
4.378 &
10.510 &
17.206\tabularnewline
\texttt{nloptr} &
64.904 &
117.533 &
115.738 &
226.661\tabularnewline
\texttt{CVXR} &
31.241 &
43.909 &
42.435 &
136.007\tabularnewline
\texttt{CVX} &
41.441 &
54.095 &
65.846 &
88.982\tabularnewline
\bottomrule
\end{tabular}
\end{table}

To evaluate the computational cost, we record the time spent in the
inner loop. With $100$ sets of data generated by the same DGP for
each sample size, we fix $\beta=\left(0.9,\,0.9\right)$ and only
numerically solve the inner loop. Since four approaches have virtually
identical inner loop results, we only report the running time of each
method in Table \ref{tab:rel_running_time}. Although \texttt{CVXR}
and \texttt{nloptr} are able to correctly solve the problem thanks
to its small scale, \texttt{Rmosek} remains $4$ to $30$ times faster
than these alternatives. We conjecture that bigger speed gain would
be observed in a problem of larger scale.

\section{Conclusion\label{sec:Conclusion}}

In this note, we demonstrate numerical implementation via \texttt{Rmosek}
of two examples of high-dimensional econometric estimators. The convenience
and reliability of high-dimensional convex optimization in \texttt{R}
will open new possibilities to create estimation procedures. In the
era of big data, we are looking forward to witnessing more algorithms
blossoming and flourishing along with theoretical research of high-dimensional
models.

\bigskip
\singlespacing\bibliographystyle{chicagoa}
\bibliography{sample}
\onehalfspacing

\newpage{}

\appendix
\begin{center}
\textbf{\huge{}Appendix}{\huge\par}
\par\end{center}

\begin{center}
{\Large{}(To be published online only)}{\Large\par}
\par\end{center}

\section{Data Generating Process}

For completeness of the note, in this section we detail the DGPs and
simulation design.

\subsection{C-Lasso\label{subsec:C-Lasso DGP}}

We follow the linear static panel data DGP (DGP 1) in \citet[p.2237]{su2016identifying}
and apply PLS. The observations are drawn from three groups with the
proportion $n_{1}:n_{2}:n_{3}=0.3:0.3:0.4$. The observed data $\left(y_{it},x_{it}\right)$
are generated from 
\begin{align*}
x_{it} & =\left(0.2\mu_{i}^{0}+e_{it1},0.2\mu_{i}^{0}+e_{it2}\right)'\\
y_{it} & =\beta_{i}^{0\prime}x_{it}+\mu_{i}^{0}+\varepsilon_{it},
\end{align*}
where $\mu_{i}^{0}$, $\varepsilon_{it}$, $e_{it1}$, $e_{it2}\sim\text{i.i.d.}N\left(0,1\right)$.
The true coefficients are $\left(0.4,1.6\right)$, $\left(1,1\right)$,
$\left(1.6,0.4\right)$ for the three groups, respectively. In the
implementation, the C-Lasso tuning parameter is specified as $\lambda=\frac{1}{2}\widehat{\sigma}{}_{Y}^{2}T^{-\frac{1}{3}}$,
where $\widehat{\sigma}_{Y}^{2}$ is the sample variance of demeaned
dependent variable. Given the number of groups, we run the simulation
for $R=500$ replications and report the RMSE of the estimation of
$\alpha_{1}$ and the probability of correct classification (correct
ratio) in Table \ref{tab:classopls}, where
\begin{eqnarray*}
\mbox{RMSE}\left(\hat{\beta}_{1}\right) & = & \sqrt{\frac{1}{R}\sum_{r=1}^{R}\left(\sum_{k=1}^{K}\frac{n_{k}}{n}\left(\hat{\alpha}_{k,1}^{\left(r\right)}-\alpha_{k,1}^{0}\right)^{2}\right)}\\
\mbox{Correct Ratio} & = & \frac{1}{R}\sum_{r=1}^{R}\left(\frac{1}{n}\sum_{i=1}^{n}\boldsymbol{1}\left(\hat{g}_{i}^{\left(r\right)}=g_{i}^{0}\right)\right),
\end{eqnarray*}
where $\widehat{g}_{i}^{\left(r\right)}$ and $g_{i}^{\left(0\right)}$
are the estimated and the true group identity of the $i$'s individual,
respectively, and $\boldsymbol{1}\left(\cdot\right)$ is the indicator
function.

\subsection{REL\label{subsec:REL-DGP}}

We follow the data generating process in \citet[Section 4.1]{REL}
that features the linear IV model with many IVs. The observed data
$\left\{ y_{i}\right\} _{i=1}^{n}$ are generated by the structural
equation
\[
y_{i}=\left(x_{i1},x_{i2}\right)\beta+e_{i}^{0}
\]
where $\beta=\left(1,1\right)^{\prime}$, $x_{i}=\left(x_{i1},x_{i2}\right)$
are endogenous variables that are generated by $x_{i1}=0.5z_{i1}+0.5z_{i2}+e_{i}^{1}$
and $x_{i2}=0.5z_{i3}+0.5z_{i4}+e_{i}^{2}$, respectively, $e_{i}^{0}$
is the structural error, and $\left(e_{i}^{1},e_{i}^{2}\right)$ are
reduced-form errors. The observed data contains $m$ IVs $\left\{ z_{ij}\right\} _{j=1}^{m}$
orthogonal to $e_{i}^{0}$ but the information that which one is relevant
is unknown. We generate $\left\{ z_{ij}\right\} _{j=1}^{m}\sim\text{i.i.d.}n\left(0,1\right)$
and $\begin{pmatrix}e_{i}^{0}\\
e_{i}^{1}\\
e_{i}^{2}
\end{pmatrix}\sim n\left(\begin{pmatrix}0\\
0\\
0
\end{pmatrix},\begin{pmatrix}0.25 & 0.15 & 0.15\\
0.15 & 0.25 & 0\\
0.15 & 0 & 0.25
\end{pmatrix}\right)$. The endogeneity comes from the correlation among all error terms.
The orthogonality yields the moment restrictions $\mathbb{E}\left[z_{i}\left(y_{i}-x_{i}\beta\right)\right]=\boldsymbol{0}_{m}$
which can be used to estimate $\beta$ with REL. We run $R=500$ replications
and report bias and RMSE for $\beta_{1}$ as $\text{Bias}=\frac{1}{R}\sum_{r=1}^{R}\left(\hat{\beta}_{1}-\beta_{1}\right)$
and $\text{RMSE}=\sqrt{\frac{1}{R}\sum_{r=1}^{R}\left(\hat{\beta}_{1}-\beta_{1}\right)^{2}}$.

\section{Code Snippets\label{sec:Code-Snippets}}

In this section, we provide several code snippets to demonstrate the
key formulation steps. All code in this note is hosted at \url{https://github.com/zhentaoshi/convex_prog_in_econometrics}. 

We start with Lasso. In matrix notation, the Lasso problem is 
\begin{gather*}
\min_{\theta}\,\lambda\left(e'\beta^{+}+e'\beta^{-}\right)+\frac{t}{n}\quad\\
\text{s.t.}\ \ \begin{bmatrix}\begin{array}{ccc}
X & -X & I_{n}\end{array} & \boldsymbol{0}_{n\times3}\\
\boldsymbol{0}_{2\times\left(n+2p\right)} & \begin{array}{ccc}
-\frac{1}{2} & 1 & 0\\
-\frac{1}{2} & 0 & 1
\end{array}
\end{bmatrix}\theta=\begin{bmatrix}y\\
-\frac{1}{2}\\
\frac{1}{2}
\end{bmatrix},\,\lVert(v,s)\rVert_{2}\leq r,\,\beta^{+},\beta^{-}\geq0
\end{gather*}
where the inequality for a vector is taken elementwisely. The following
annotated \texttt{R} code snippet implements the matrix form.

\begin{knitrout}
\definecolor{shadecolor}{rgb}{0.969, 0.969, 0.969}\color{fgcolor}\begin{kframe}
\begin{alltt}
\hlstd{P} \hlkwb{=} \hlkwd{list}\hlstd{(}\hlkwc{sense} \hlstd{=} \hlstr{"min"}\hlstd{)}

\hlcom{# Linear coefficients in objective}
\hlstd{P}\hlopt{$}\hlstd{c} \hlkwb{=} \hlkwd{c}\hlstd{(}\hlkwd{rep}\hlstd{(lambda,} \hlnum{2}\hlopt{*}\hlstd{p),} \hlkwd{rep}\hlstd{(}\hlnum{0}\hlstd{, n),} \hlnum{1}\hlopt{/}\hlstd{n,} \hlnum{0}\hlstd{,} \hlnum{0}\hlstd{)}

\hlcom{# The matrix in linear constraints}
\hlstd{A} \hlkwb{=} \hlkwd{as.matrix.csr}\hlstd{(X)}
\hlstd{A} \hlkwb{=} \hlkwd{cbind}\hlstd{(A,} \hlopt{-}\hlstd{A,} \hlkwd{as}\hlstd{(n,} \hlstr{"matrix.diag.csr"}\hlstd{),} \hlkwd{as.matrix.csr}\hlstd{(}\hlnum{0}\hlstd{, n,} \hlnum{3}\hlstd{))}
\hlstd{A} \hlkwb{=} \hlkwd{rbind}\hlstd{(A,} \hlkwd{cbind}\hlstd{(}\hlkwd{as.matrix.csr}\hlstd{(}\hlnum{0}\hlstd{,} \hlnum{2}\hlstd{,} \hlnum{2}\hlopt{*}\hlstd{p} \hlopt{+} \hlstd{n),}
                                 \hlkwd{as.matrix.csr}\hlstd{(}\hlkwd{c}\hlstd{(}\hlopt{-}\hlnum{.5}\hlstd{,} \hlopt{-}\hlnum{.5}\hlstd{,} \hlnum{1}\hlstd{,} \hlnum{0}\hlstd{,} \hlnum{0}\hlstd{,} \hlnum{1}\hlstd{),} \hlnum{2}\hlstd{,} \hlnum{3}\hlstd{)))}
\hlstd{P}\hlopt{$}\hlstd{A} \hlkwb{=} \hlkwd{as}\hlstd{(A,}\hlstr{"CsparseMatrix"}\hlstd{)}

\hlcom{# Right-hand side of linear constraints}
\hlstd{P}\hlopt{$}\hlstd{bc} \hlkwb{=} \hlkwd{rbind}\hlstd{(}\hlkwd{c}\hlstd{(y,} \hlopt{-}\hlnum{0.5}\hlstd{,} \hlnum{0.5}\hlstd{),} \hlkwd{c}\hlstd{(y,} \hlopt{-}\hlnum{0.5}\hlstd{,} \hlnum{0.5}\hlstd{))}

\hlcom{# Constraints on variables}
\hlstd{P}\hlopt{$}\hlstd{bx} \hlkwb{=} \hlkwd{rbind}\hlstd{(}\hlkwd{c}\hlstd{(}\hlkwd{rep}\hlstd{(}\hlnum{0}\hlstd{,} \hlnum{2} \hlopt{*} \hlstd{p),} \hlkwd{rep}\hlstd{(}\hlopt{-}\hlnum{Inf}\hlstd{, n),} \hlkwd{rep}\hlstd{(}\hlnum{0}\hlstd{,} \hlnum{3}\hlstd{)),} \hlkwd{c}\hlstd{(}\hlkwd{rep}\hlstd{(}\hlnum{Inf}\hlstd{,} \hlnum{2}\hlopt{*}\hlstd{p}\hlopt{+}\hlstd{n}\hlopt{+}\hlnum{3}\hlstd{)))}

\hlcom{# Conic constraints}
\hlstd{P}\hlopt{$}\hlstd{cones} \hlkwb{=} \hlkwd{matrix}\hlstd{(}\hlkwd{list}\hlstd{(}\hlstr{"QUAD"}\hlstd{,} \hlkwd{c}\hlstd{(n}\hlopt{+}\hlnum{2}\hlopt{*}\hlstd{p}\hlopt{+}\hlnum{3}\hlstd{, (}\hlnum{2}\hlopt{*}\hlstd{p}\hlopt{+}\hlnum{1}\hlstd{)}\hlopt{:}\hlstd{(}\hlnum{2}\hlopt{*}\hlstd{p}\hlopt{+}\hlstd{n), n}\hlopt{+}\hlnum{2}\hlopt{*}\hlstd{p}\hlopt{+}\hlnum{2}\hlstd{)),} \hlnum{2}\hlstd{,} \hlnum{1}\hlstd{)}
\hlkwd{rownames}\hlstd{(P}\hlopt{$}\hlstd{cones)} \hlkwb{=} \hlkwd{c}\hlstd{(}\hlstr{"type"}\hlstd{,} \hlstr{"sub"}\hlstd{)}

\hlstd{result} \hlkwb{=} \hlkwd{mosek}\hlstd{(P,} \hlkwc{opts} \hlstd{=} \hlkwd{list}\hlstd{(}\hlkwc{verbose} \hlstd{= verb))}
\hlstd{xx} \hlkwb{=} \hlstd{result}\hlopt{$}\hlstd{sol}\hlopt{$}\hlstd{itr}\hlopt{$}\hlstd{xx}
\hlstd{coef} \hlkwb{=} \hlstd{xx[}\hlnum{1}\hlopt{:}\hlstd{p]} \hlopt{-} \hlstd{xx[(p}\hlopt{+}\hlnum{1}\hlstd{)}\hlopt{:}\hlstd{(}\hlnum{2}\hlopt{*}\hlstd{p)]}
\end{alltt}
\end{kframe}
\end{knitrout}

We then take a step further to C-Lasso. The convexity is manifest
when we write the problem in matrix form 
\begin{gather*}
\min_{\alpha_{\tilde{k}},\theta}\,\left(\frac{1}{nT}\right)e^{\prime}t+\left(\frac{\lambda}{n}\right)\gamma^{\prime}w\\
\text{s.t. \ }\ t_{i}\geq0,\,\left\Vert \left(\nu_{i},s_{i}\right)\right\Vert _{2}\leq r_{i},\,\left\Vert \mu_{i}\right\Vert _{2}\leq w_{i},\,\text{for all\,}i=1,2,\cdots,n\\
\begin{bmatrix}\begin{array}{ccc}
\mathrm{diag}\left(X_{1},\ldots,X_{n}\right) & I_{Tn} & \boldsymbol{0}\\
I_{np} & \boldsymbol{0} & \begin{array}{c}
-I_{np}\end{array}
\end{array} & \boldsymbol{0} & \begin{array}{c}
\boldsymbol{0}\\
-\boldsymbol{1}_{n}\otimes I_{p}
\end{array}\\
\boldsymbol{0} & \begin{array}{cc}
I_{2}\otimes I_{n} & -\frac{1}{2}\boldsymbol{1}_{2}\otimes I_{n}\end{array} & \boldsymbol{0}
\end{bmatrix}\begin{bmatrix}\theta\\
\alpha_{\tilde{k}}
\end{bmatrix}=\begin{bmatrix}y\\
\boldsymbol{0}_{np}\\
-\frac{1}{2}e_{n}\\
\frac{1}{2}e_{n}
\end{bmatrix}
\end{gather*}

Though more tedious than Lasso, the construction of the large matrix
in the linear constraints is straightforward. The formulation of the
conic constraints is illustrated in the following chunk of code.

\begin{knitrout}
\definecolor{shadecolor}{rgb}{0.969, 0.969, 0.969}\color{fgcolor}\begin{kframe}
\begin{alltt}
\hlstd{CC} \hlkwb{=} \hlkwd{list}\hlstd{()}

\hlcom{# locate the variables related}
\hlstd{bench} \hlkwb{=} \hlstd{N}\hlopt{*}\hlstd{(}\hlnum{2}\hlopt{*}\hlstd{p} \hlopt{+} \hlstd{TT)} \hlopt{+} \hlstd{p}

\hlkwa{for}\hlstd{(i} \hlkwa{in} \hlnum{1}\hlopt{:}\hlstd{N)\{}
        \hlcom{# find index of each variable}
        \hlstd{s.i} \hlkwb{=} \hlstd{bench} \hlopt{+} \hlstd{i}
        \hlstd{r.i} \hlkwb{=} \hlstd{bench} \hlopt{+} \hlstd{N} \hlopt{+} \hlstd{i}
        \hlstd{nu.i} \hlkwb{=} \hlstd{(N}\hlopt{*}\hlstd{p} \hlopt{+} \hlstd{(i}\hlopt{-}\hlnum{1}\hlstd{)}\hlopt{*}\hlstd{TT} \hlopt{+} \hlnum{1}\hlstd{)}\hlopt{:}\hlstd{(N}\hlopt{*}\hlstd{p} \hlopt{+} \hlstd{i}\hlopt{*}\hlstd{TT)}
        \hlstd{w.i} \hlkwb{=} \hlstd{bench} \hlopt{+} \hlnum{3}\hlopt{*}\hlstd{N} \hlopt{+} \hlstd{i}
        \hlstd{mu.i} \hlkwb{=} \hlstd{(N}\hlopt{*}\hlstd{(TT}\hlopt{+}\hlstd{p)} \hlopt{+} \hlstd{(i}\hlopt{-}\hlnum{1}\hlstd{)}\hlopt{*}\hlstd{p} \hlopt{+} \hlnum{1}\hlstd{)}\hlopt{:}\hlstd{(N}\hlopt{*}\hlstd{(TT}\hlopt{+}\hlstd{p)} \hlopt{+} \hlstd{i}\hlopt{*}\hlstd{p)}
        \hlstd{CC} \hlkwb{=} \hlkwd{cbind}\hlstd{(CC,} \hlkwd{list}\hlstd{(}\hlstr{"QUAD"}\hlstd{,} \hlkwd{c}\hlstd{(r.i, nu.i, s.i)),}
                                   \hlkwd{list}\hlstd{(}\hlstr{"QUAD"}\hlstd{,} \hlkwd{c}\hlstd{(w.i, mu.i)) )}
\hlstd{\}}
\hlstd{P}\hlopt{$}\hlstd{cones} \hlkwb{=} \hlstd{CC}
\hlkwd{rownames}\hlstd{(prob}\hlopt{$}\hlstd{cones)} \hlkwb{=} \hlkwd{c}\hlstd{(}\hlstr{"type"}\hlstd{,} \hlstr{"sub"}\hlstd{)}
\end{alltt}
\end{kframe}
\end{knitrout}

The penalty $\gamma_{i}$ can be coded as follows.

\begin{knitrout}
\definecolor{shadecolor}{rgb}{0.969, 0.969, 0.969}\color{fgcolor}\begin{kframe}
\begin{alltt}
\hlstd{pen.generate} \hlkwb{=} \hlkwa{function}\hlstd{(}\hlkwc{b}\hlstd{,} \hlkwc{a}\hlstd{,} \hlkwc{N}\hlstd{,} \hlkwc{p}\hlstd{,} \hlkwc{K}\hlstd{,} \hlkwc{kk}\hlstd{)\{}

        \hlcom{# Output arg: gamma}
        \hlcom{# Input args: }
        \hlcom{#	 b, a (estimate of last iteration)}
        \hlcom{# 	kk (current focused group)  }

        \hlcom{# compute all ||\textbackslash{}beta_i - alpha_k||_2}
        \hlstd{a.out.exp} \hlkwb{=} \hlkwd{aperm}\hlstd{(}\hlkwd{array}\hlstd{(a,} \hlkwd{c}\hlstd{(K, p, N)),} \hlkwd{c}\hlstd{(}\hlnum{3}\hlstd{,} \hlnum{2}\hlstd{,} \hlnum{1}\hlstd{))}
        \hlstd{p.norm} \hlkwb{=} \hlkwd{sqrt}\hlstd{(}\hlkwd{apply}\hlstd{((b} \hlopt{-} \hlstd{a.out.exp)}\hlopt{^}\hlnum{2}\hlstd{,} \hlkwd{c}\hlstd{(}\hlnum{1}\hlstd{,}\hlnum{3}\hlstd{), sum))}

        \hlcom{# leave kk out and take product}
        \hlstd{ind} \hlkwb{=} \hlkwd{setdiff}\hlstd{(}\hlnum{1}\hlopt{:}\hlstd{K,kk)}
        \hlstd{gamma} \hlkwb{=} \hlkwd{apply}\hlstd{(p.norm[, ind],} \hlnum{1}\hlstd{, prod)}
        \hlkwd{return}\hlstd{(gamma)}
\hlstd{\}}
\end{alltt}
\end{kframe}
\end{knitrout}

Regarding REL, it involves nonlinear logarithm terms in the objective.
The objective of the separable convex problem can be formulated as
follows.

\begin{knitrout}
\definecolor{shadecolor}{rgb}{0.969, 0.969, 0.969}\color{fgcolor}\begin{kframe}
\begin{alltt}
\hlstd{NUMOPRO} \hlkwb{=} \hlstd{n}
\hlstd{opro} \hlkwb{=} \hlkwd{matrix}\hlstd{(}\hlkwd{list}\hlstd{(),} \hlkwc{nrow} \hlstd{=} \hlnum{5}\hlstd{,} \hlkwc{ncol} \hlstd{= NUMOPRO)}
\hlkwd{rownames}\hlstd{(opro)} \hlkwb{=} \hlkwd{c}\hlstd{(}\hlstr{"type"}\hlstd{,} \hlstr{"j"} \hlstd{,} \hlstr{"f"}\hlstd{,} \hlstr{"g"}\hlstd{,} \hlstr{"h"}\hlstd{)}
\hlkwa{for}\hlstd{(i} \hlkwa{in} \hlnum{1}\hlopt{:}\hlstd{n)\{}
        \hlstd{opro[ , i]} \hlkwb{=} \hlkwd{list}\hlstd{(}\hlstr{"LOG"}\hlstd{, i,} \hlnum{1.0}\hlstd{,} \hlnum{1.0}\hlstd{,} \hlnum{0}\hlstd{)}
\hlstd{\}}
\hlstd{P}\hlopt{$}\hlstd{scopt} \hlkwb{=} \hlkwd{list}\hlstd{(}\hlkwc{opro} \hlstd{= opro)}
\end{alltt}
\end{kframe}
\end{knitrout}

\section{Additional Examples of C-Lasso}

In this section, we formulate the nonlinear Lasso and the penalized
GMM (PGMM).

\subsection{Nonlinear Lasso\label{subsec:Nonlinear-Lasso}}

In microeconometrics, it is common to see exponential, logarithm or
power terms in objective functions. When the problem involves these
nonlinear functions, we formulate the problem as a separable convex
optimization problem. For example, the penalized Poison maximum likelihood
estimator is defined as 
\[
\min_{\beta}\,-\frac{1}{n}\sum_{i=1}^{n}\left(y_{i}x_{i}'\beta-\exp\left(x_{i}'\beta\right)\right)+\lambda\lVert\beta\rVert_{1}
\]
where $y\in\mathbb{R}^{n}$ and $X\in\mathbb{R}^{n\times p}$ are
observed data, $\lambda$ is the tuning parameter and $\beta\in\mathbb{R}^{p}$
is the parameter of interests. This optimization problem involves
the component $\exp\left(\sum_{j=1}^{p}x_{ij}\beta_{j}\right)$, which
is non-separable. Define $v_{i}=x_{i}'\beta$, and the objective becomes
\[
\min_{v,\beta}\,-\frac{1}{n}\sum_{i=1}^{n}\left(y_{i}v_{i}-\exp\left(v_{i}\right)\right)+\lambda\lVert\beta\rVert_{1}
\]
We apply the same transformation as in Lasso to deal with the $l_{1}$-norm.
The original optimization problem can be transformed to
\begin{gather*}
\min_{v,\beta^{+},\beta^{-}}\,-\frac{1}{n}\sum_{i=1}^{n}\left(y_{i}v_{i}-\exp\left(v_{i}\right)\right)+\lambda\left(e'\beta^{+}+e'\beta^{-}\right)\\
\text{s.t.}\ \ v_{i}=x_{i}'\left(\beta^{+}-\beta^{-}\right)\:\text{for each}\,i=1,2,3,\cdots,n,\,\beta^{+},\beta^{-}\geq0
\end{gather*}
In matrix form,
\begin{gather*}
\min_{\theta}\,\begin{bmatrix}-y^{\prime} & \lambda e^{\prime} & \lambda e^{\prime}\end{bmatrix}\theta+\frac{1}{n}\sum_{i=1}^{n}\exp(v_{i})\\
\text{s.t.}\ \ \begin{bmatrix}I_{n} & -X & X\end{bmatrix}\theta=0,\,\beta^{+},\beta^{-}\geq0
\end{gather*}
where $\theta=\left(v,\beta^{+},\beta^{-}\right).$ The following
code snippet displays the formulation of these exponential terms.

\begin{knitrout}
\definecolor{shadecolor}{rgb}{0.969, 0.969, 0.969}\color{fgcolor}\begin{kframe}
\begin{alltt}
\hlstd{NUMOPRO} \hlkwb{=} \hlstd{n}
\hlstd{opro} \hlkwb{=} \hlkwd{matrix}\hlstd{(}\hlkwd{list}\hlstd{(),} \hlkwc{nrow} \hlstd{=} \hlnum{5}\hlstd{,} \hlkwc{ncol} \hlstd{= NUMOPRO)}
\hlkwd{rownames}\hlstd{(opro)} \hlkwb{=} \hlkwd{c}\hlstd{(}\hlstr{"type"}\hlstd{,} \hlstr{"j"} \hlstd{,} \hlstr{"f"}\hlstd{,} \hlstr{"g"}\hlstd{,} \hlstr{"h"}\hlstd{)}
\hlkwa{for}\hlstd{(i} \hlkwa{in} \hlnum{1}\hlopt{:}\hlstd{n)\{}
        \hlstd{opro[,i]} \hlkwb{=} \hlkwd{list}\hlstd{(}\hlstr{"EXP"}\hlstd{, i,} \hlnum{1}\hlopt{/}\hlstd{n,} \hlnum{1.0}\hlstd{,} \hlnum{0}\hlstd{)}
\hlstd{\}}
\hlstd{P}\hlopt{$}\hlstd{scopt} \hlkwb{=} \hlkwd{list}\hlstd{(}\hlkwc{opro}\hlstd{=opro)}
\end{alltt}
\end{kframe}
\end{knitrout}

Now that we are able to deal with nonlinear Lasso, it is straightforward
to extend it to penalized profile likelihood (PPL) in \citet{su2016identifying}.

\subsection{Penalized GMM\label{subsec:Penalized-GMM}}

We consider the linear panel data model with latent group structures
and endogeneity. After first-difference, we have 
\[
\Delta y_{it}=\beta_{i}^{\prime}\Delta x_{it}+\Delta\varepsilon_{it}
\]
Let $z_{it}$, of dimension $m\times1$, $m\geq p$, be instrumental
variables for $\Delta x_{it}$. The penalized GMM estimator is defined
as the solution $\left(\boldsymbol{\beta},\alpha\right)$ to
\[
\min_{\boldsymbol{\beta},\alpha}\:\frac{1}{nT^{2}}\sum_{i=1}^{n}\left\Vert W_{i}^{\frac{1}{2}}z_{i}\left(\Delta y_{i}-\Delta x_{i}\beta_{i}\right)\right\Vert _{2}^{2}+\frac{\lambda}{n}\sum_{i=1}^{n}\prod_{k=1}^{K}\left\Vert \beta_{i}-\alpha_{k}\right\Vert _{2}
\]
where $W_{i}$ is an $m\times m$ positive-definite symmetric weighting
matrix. It is easy to see that the PGMM problem can be formulated
as
\[
\min_{\boldsymbol{\beta},\alpha}\,\frac{1}{nT^{2}}\sum_{i=1}^{n}\left\Vert \tilde{y}_{i}-\tilde{x}_{i}\beta_{i}\right\Vert _{2}^{2}+\frac{\lambda}{n}\sum_{i=1}^{n}\prod_{k=1}^{K}\left\Vert \beta_{i}-\alpha_{k}\right\Vert _{2}
\]
by the transformations $\tilde{y}_{i}=W_{i}^{\frac{1}{2}}z_{i}\Delta y_{i}$
and $\tilde{x}_{i}=W_{i}^{\frac{1}{2}}z_{i}\Delta x_{i}$. The following
iterative algorithm is essentially the same as PLS and can be carried
out as in Section \ref{sec:Classifier-Lasso}.

\section{Software Installation\label{sec:Computing-Environment}}

The installation of the \texttt{Rmosek} package requires successful
installation of \texttt{MOSEK}. For Windows users, \texttt{Rtools}
is also required. The installation of the latest version \texttt{MOSEK}
9.0 includes \texttt{Rmosek} and it can be invoked in \texttt{R}:

Once the prerequisites are satisfied, \texttt{Rmosek} can be installed
by a command similar to the following one:

\begin{knitrout}
\definecolor{shadecolor}{rgb}{0.969, 0.969, 0.969}\color{fgcolor}\begin{kframe}
\begin{alltt}
\hlkwd{source}\hlstd{(}\hlstr{"<RMOSEKDIR>/builder.R"}\hlstd{)}
\hlkwd{attachbuilder}\hlstd{()}
\hlkwd{install.rmosek}\hlstd{()}
\end{alltt}
\end{kframe}
\end{knitrout}

For more details, readers can refer to the official installation manual
at \url{https://docs.mosek.com/9.0/rmosek/install-interface.html}.

\texttt{CVXR} is now available on CRAN and can be installed as a standard
\texttt{R} packages. The default solver \texttt{ECOS} is installed
along with \texttt{CVXR}. To use \texttt{MOSEK} in \texttt{CVXR},
we will need \texttt{Python} and the \texttt{R} package \texttt{reticulate}.
Details can be found at \url{https://cvxr.rbind.io/cvxr_examples/cvxr_using-other-solvers/}.
\end{document}